%% file: main.tex
\documentclass[a4paper, oneside, twocolumn, notitlepage, 10pt]{extarticle_ecoc}
\usepackage{ecoc}
\pdfoutput=1

\newcommand{\Lawgn}{L_k^\mathrm{awgn}}
\newcommand{\Lzc}{L_k^\mathrm{zc}}

\begin{document}
\selectlanguage{english}    % Standard Language
\pdfoutput=1

%-------------------------------------------------- Title -----------------------------------------------------%

\title{On LLR Calculations for Soft-decision Decoding in Next-generation IM-DD Systems with Laser RIN}%

%------------------------------------------------- Authors-----------------------------------------------------%

\author{
    %\textsuperscript{(1)}
    Felipe Villenas\textsuperscript{(*)}, Yunus Can G\"ultekin, and Alex Alvarado
}

\maketitle                  % Create title and author

%------------------------------------------ Description of Authors ----------------------------------------------%

\begin{strip}
    \begin{author_descr}
    
        %Information and Communication Theory Lab, Eindhoven University of Technology, the Netherlands, \textsuperscript{(*)}\textcolor{blue}{\uline{f.i.villenas.cortez@tue.nl}}
        Deparment of Electrical Engineering, Eindhoven University of Technology, Eindhoven, the Netherlands, \textsuperscript{(*)}\textcolor{blue}{\uline{f.i.villenas.cortez@tue.nl}}
        
    \end{author_descr}
\end{strip}

% \setstretch{1.1}
%-------------------------------------------------- Footnote -------------------------------------------------------%
\renewcommand\footnotemark{}
\renewcommand\footnoterule{}
%\let\thefootnote\relax\footnotetext{text}

%-------------------------------------------------- Abstract ---------------------------------------------------------%

\begin{strip}
    \begin{ecoc_abstract}
        % NOTE: Don't use a blank line here but start abstract right away to avoid an extra line break
        %We propose a low-complexity LLR approximation for SD decoding in IM-DD systems with relative intensity noise. Our approximation has no BER penalty, and we show that, for a concatenated KP4+Hamming FEC scheme, it outperforms mismatched AWGN-based LLRs. ©2026 The Author(s)
        We propose a low-complexity LLR approximation for SD decoding in IM-DD systems with relative intensity noise. Our approximation results in no BER performance penalty for a concatenated KP4+Hamming FEC scheme, and outperforms mismatched AWGN-based LLRs. ©2026 The Author(s)
    \end{ecoc_abstract}
\end{strip}

%-------------------------------------------------- Introduction Section -------------------------------------------------------%

\section{Introduction}

\begin{figure*}[!b]
    \centering
    \input{fig/combined_system.tikz}
    \caption{Considered system: concatenated KP4+Hamming FEC scheme with an IM-DD channel with signal-dependent noise. The eye diagram shows a PAM-4 signal after the ADC, where each PAM level has a different noise distribution due to RIN as shown by the red Gaussians. The LLRs $L_k$ are calculated with either \eqref{eq:Lk_2}, \eqref{eq:Lk_awgn}, or \eqref{eq:L1_zc} and \eqref{eq:L2_zc}, and then fed to the SD decoder.}
    \label{fig:system}
\end{figure*}
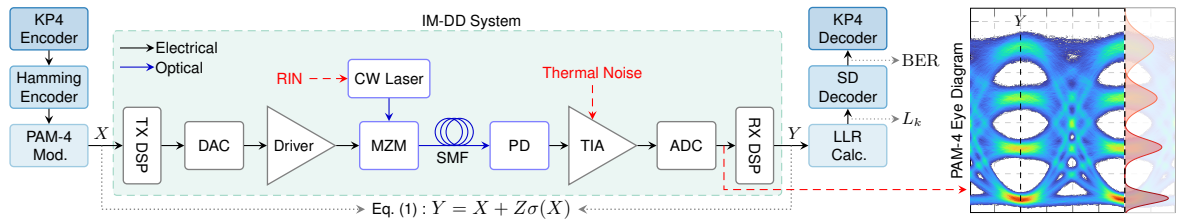

The rapid growth of artificial intelligence applications is driving demand for higher speeds in short-reach data center intraconnects (DCI) \cite{siamak2026}. Low hardware costs and low power consumption are essential, and thus, DCI optical transceivers employ intensity-modulation (IM) and direct-detection (DD) with 4-ary pulse amplitude modulation (PAM-4) \cite{zhou2026}. Scaling the data rates to 400 Gb/s/lane and beyond faces significant challenges due to the bandwidth constraints of electro-optical components and noise impairments \cite{pang2020200,tatarczak2026}. One important noise impairment in such systems is relative intensity noise (RIN) arising from the laser source.

To achieve the scaling to 400 Gb/s per lane and beyond, several technologies have been studied, such as advanced digital signal processing (DSP) techniques \cite{oettinghaus2026advanced} or forward error correction (FEC) codes with low code rates \cite{wang2023advancedFEC}. To improve the coding gain with respect to KP4, several proposals have been made. For example, a concatenated FEC scheme based on the standard KP4 code \cite{baseline_200G} as the outer code with hard-decision (HD) decoding, and a Hamming code as the inner code with \emph{soft-decision} (SD) \emph{decoding} was proposed for the 200 Gb/s/lane baseline in \cite{farhood_P802_3dj}. 

SD decoding typically uses log-likelihood ratios (LLR). Calculating LLRs requires knowledge of the channel, which is not always straightforward to obtain accurately \cite{bosco2003soft}. Channels with signal-dependent noise, such as RIN, induce different noise statistics per symbol. LLR calculations for PAM systems with signal-dependent noise have been studied in works such as \cite{shi2025enhanced}, \cite{zhou2020improved} (using machine learning), and \cite{pan2023adaptive} (using adaptive calculation algorithms). For the additive white Gaussian noise (AWGN) channel, LLRs are well approximated by piecewise linear functions of the received values. However, for channels with signal-dependent noise such as RIN, this is no longer the case. Both LLR expressions and low-complexity approximations for PAM-4 in high-speed DCI IM-DD systems with RIN have yet to be presented.

In this paper, we study the LLR calculations of an IM-DD system limited by RIN. First, we show that the LLRs for this channel follow a quadratic polynomial relationship as a function of the received values. We also show that using mismatched AWGN-based LLRs that neglect signal-dependent noise results in a significant penalty in terms of post-FEC bit error rate (BER). Furthermore, we propose a low-complexity piecewise linear approximation of the LLRs and show that there is virtually no post-FEC BER performance loss when using our proposed approximation as the soft information for the SD decoder.

\section{System Model}
We consider the concatenated FEC scheme depicted in Fig.~\ref{fig:system}. KP4-coded bits are encoded with a $(128, 120)$ extended Hamming code, and then modulated into PAM-4 symbols $X\in\mathcal{X}=\{\pm\Delta,\pm3\Delta\}$ using Gray labeling, where $\Delta$ is a scaling factor. The symbols $X$ are the input to the standard IM-DD system shown in Fig.~\ref{fig:system}. The transmitter (TX) DSP applies pulse shaping and nonlinear pre-distortion to compensate for the modulator transfer function. The DAC generates the analog waveform which is then conditioned by the driver for IM. A Mach-Zehnder modulator (MZM) is used to modulate the intensity of an O-band ($1310$ nm) continuous-wave (CW) laser with RIN.

The optical signal is transmitted through a single-mode fiber (SMF). Given the short fiber length of $<\!500$ m, both chromatic dispersion and attenuation are negligible. At the receiver (RX), a photodiode (PD) generates current proportional to the incident optical light. The signal bias from IM is removed and a transimpedance amplifier (TIA) converts the current to voltage. During this process, thermal noise is generated. 

The analog waveform is sampled by the ADC to be further processed by the receiver DSP. The RX DSP applies matched filtering, downsampling, and equalization to remove intersymbol interference (ISI). For this work, we consider any remaining ISI after equalization to be negligible. Fig.~\ref{fig:system} shows an example of an eye diagram for the system under consideration, where the presence of RIN is clearly visible. RIN generates a noise distribution proportional to the square of the symbol's optical amplitude, as depicted by the red Gaussians at the maximum eye opening. 

We consider the same channel model as in \cite{villenas2025ofc,villenas2025ecoc}, where the channel output $Y$ is modeled as
\begin{equation} \label{eq:ch_model}
    Y = X+Z\sigma(X),
\end{equation}
where $Z$ is a zero-mean, and unit-variance Gaussian random variable to model the noise. The total noise variance is modeled with the function $\sigma(X)$, defined as
\begin{equation} \label{eq:noise_model}
    \sigma^2(X)\triangleq p_0+p_1(X+\beta)^2,
\end{equation}
where $p_0$ and $p_1$ are coefficients that depend on thermal noise and RIN noise statistics, respectively \cite[Eq.~(10)]{szczerba20124}, \cite[Eq.~(8)]{rizzelli2023analytical}. Furthermore, the IM bias $\beta=3\Delta\frac{\mathrm{ER}+1}{\mathrm{ER}-1}\geq |\min\{\mathcal{X}\}|$ models the extinction ratio $\mathrm{ER}$ of the system \cite[Sec.~III]{che2021does}.

For the channel model in \eqref{eq:ch_model}, the conditional probability density function (PDF) of the output is
\begin{equation} \label{eq:pdf}
    f_{Y|X}(y|x) = \frac{1}{\sqrt{2\pi\sigma^2(x)}}\mathrm{exp}\left(-\frac{(y-x)^2}{2\sigma^2(x)}\right),
\end{equation}
where $\sigma^2(x)$ is the noise variance from \eqref{eq:noise_model} conditioned on each symbol $x$. The PDF \eqref{eq:pdf} is needed by the LLR calculation block (see Fig.~\ref{fig:system}). The resulting LLRs are then fed to the SD decoder to decode the Hamming code. In this work, we present results up to the output of the SD decoder. We refer to these BER values as post-FEC BER.

\section{Log-likelihood Ratio Calculations}

For the channel output $Y$ with PDF \eqref{eq:pdf}, the LLR for the bit position $k=1,2$ is calculated as
\begin{equation}
    L_k = \log \frac{\sum_{x\in\mathcal{X}_k^1} \frac{1}{\sigma(x)}\mathrm{exp}\left(-\frac{(y-x)^2}{2\sigma^2(x)}\right)}{\sum_{x\in\mathcal{X}_k^0} \frac{1}{\sigma(x)}\mathrm{exp}\left(-\frac{(y-x)^2}{2\sigma^2(x)}\right)}, \label{eq:Lk_2}
\end{equation}
where $\mathcal{X}_k^b\subset\mathcal{X}$ is the subset of symbols where the bit in position $k$ has a value of $b\in\{0,1\}$. From \eqref{eq:Lk_2}, it can be seen that $L_k$ depends on $\sigma(x)$.

% The max-log approximation \cite[Eq.~(6)]{viterbi1998intuitive} is frequently used to reduce the calculation complexity of the LLRs. The LLRs are then approximated as
% \begin{equation}
%     L_k\!\approx\! L_k^\mathrm{m\text{-}l} \!=\! \log \frac{\max_{x\in\mathcal{X}_k^1} \frac{1}{\sigma(x)}\mathrm{exp}\left(-\frac{(y-x)^2}{2\sigma^2(x)}\right)}{\max_{x\in\mathcal{X}_k^0} \frac{1}{\sigma(x)}\mathrm{exp}\left(-\frac{(y-x)^2}{2\sigma^2(x)}\right)}. \label{eq:Lk_2}
% \end{equation}
% We denote \eqref{eq:Lk_2} as the max-log LLR calculation. Figure~\ref{fig:BERR}(a) shows that 
% The max-log approximation \cite[Eq.~(6)]{viterbi1998intuitive} is frequently used to reduce the calculation complexity of the LLRs. Using the channel PDF \eqref{eq:pdf}, the max-log LLRs $L_k$ are calculated as
% \begin{align}
%     L_k &=\log\frac{\max_{x\in\mathcal{X}_k^1} \frac{1}{\sigma(x)}\mathrm{exp}\left(-\frac{(y-x)^2}{2\sigma^2(x)}\right)}{\max_{x\in\mathcal{X}_k^0} \frac{1}{\sigma(x)}\mathrm{exp}\left(-\frac{(y-x)^2}{2\sigma^2(x)}\right)}, \label{eq:Lk_2}
% \end{align}
% where $\mathcal{X}_k^b\subset\mathcal{X}$ is the subset of symbols where the bit in position $k=1,2$ has a value of $b\in\{0,1\}$. From \eqref{eq:Lk_2}, it can be seen that $L_k$ depends on $\sigma(x)$.  

A receiver that is unaware of RIN would calculate an average variance over all transmitted and received symbols. Let this variance be denoted by $\overline{\sigma}^2$. The LLRs in \eqref{eq:Lk_2} then become
\begin{equation}
    L_k^\mathrm{awgn}=\log\frac{\sum_{x\in\mathcal{X}_k^1} \mathrm{exp}\left(-\frac{(y-x)^2}{2\overline{\sigma}^2}\right)}{\sum_{x\in\mathcal{X}_k^0} \mathrm{exp}\left(-\frac{(y-x)^2}{2\overline{\sigma}^2}\right)}.\label{eq:Lk_awgn}
\end{equation}
This coincides with the LLR calculation of a linear AWGN channel, hence we denote \eqref{eq:Lk_awgn} as $\Lawgn$.

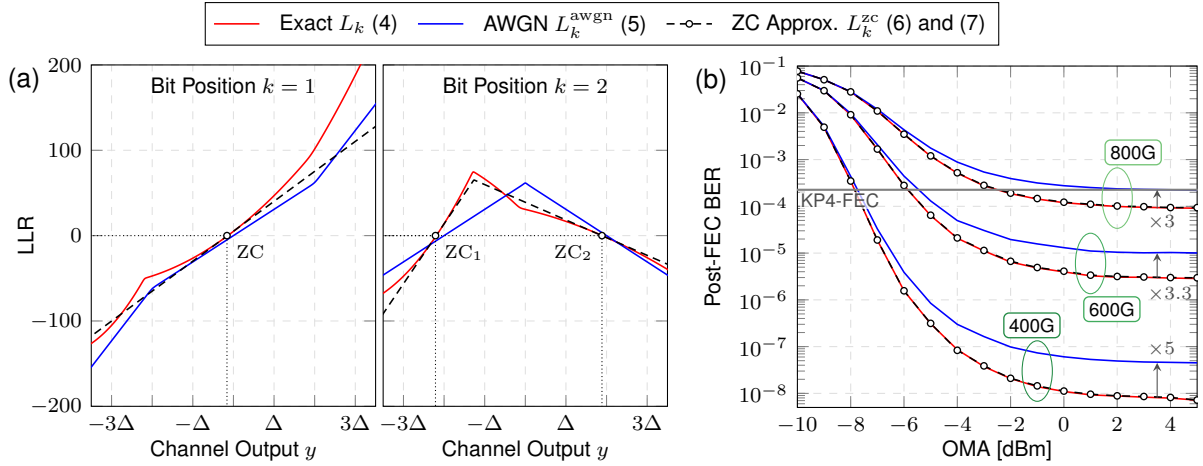
\begin{figure*}[!t]
    \centering
    \input{fig/pam4_combined}
    \caption{Comparison of the (a) LLR calculations $L_k$, $\Lawgn$, and $\Lzc$, and (b) the resulting post-FEC BER performance.}
    \label{fig:BERR}
\end{figure*}

Figure~\ref{fig:BERR}(a) shows LLR calculations for the system in Fig.~\ref{fig:system} with the parameters in Tab.~\ref{tab:sim_param}, using \eqref{eq:Lk_2} and \eqref{eq:Lk_awgn}. The results in this figure show that the exact LLRs $L_k$ behave close to a piecewise quadratic function due to the argument \mbox{$(y-x)^2/2\sigma^2(x)$} in the exponential of \eqref{eq:pdf}. On the other hand, the AWGN LLRs $\Lawgn$ behave close to a symmetric piecewise linear function of the output $y$. Furthermore, it is clear that there are noticeable differences between $L_k$ and $\Lawgn$, which is expected to lead to a performance penalty in post-FEC BER if the mismatched LLRs $\Lawgn$ are used instead of $L_k$ \cite[Sec.~V]{bosco2003soft}.

% \begin{figure}[!t]
%     \centering
%     \input{fig/pam4_llr.tikz}
%     \vspace{-1.2em}
%     \caption{Comparison of PAM-4 LLR calculations for the system in Fig.~\ref{fig:system} and parameters in Tab.~\ref{tab:sim_param}.}
%     \label{fig:LLR}
% \end{figure}

\begin{table}[!t]
    \centering
    \caption{Simulation parameters and values for 400 Gb/s}
    \begin{threeparttable}
    {\renewcommand{\arraystretch}{1.2}
    %\begin{tabular}{lclc}
    \footnotesize
    \begin{tabular}{lcl}
    \toprule
    \textbf{Parameter} & \textbf{Value} & \textbf{[Unit]} \\
    \midrule
    Symbol rate & $226$ & [GBd]\\
    Optical modulation amplitude & $0$ & [dBm]\\
    Extinction ratio & $4.5$ & [dB]\\
    %Modulation bias\tnote{*} $\beta$ & $1.05\cdot\OMA$ & [V]\\
    Link losses & $3.2$ & [dB]\\
    TIA input-referred noise & $22$ & [pA/$\sqrt{\text{Hz}}$] \\
    Relative intensity noise\hspace{10pt} & $-143$ & [dB/Hz]\\
    Noise bandwidth & $135.5$ & [GHz]\\
    Constant $\sigma^2(X):p_0$ & $1.219\cdot10^{-10}$ & [V$^2$]\\
    Constant $\sigma^2(X):p_1$ & $1.358\cdot10^{-3}$ & [$-$]\\
    \bottomrule
    \end{tabular}}
    %\begin{tablenotes}
    %\footnotesize
    %\item[*] The modulation bias is calculated as $\beta=\frac{\OMA}{2}\frac{\ER+1}{\ER-1}$.
    %\item[*]Derived from the double-sided power spectral density.
    %\end{tablenotes}
    \end{threeparttable}
    \label{tab:sim_param}
    %\vspace{-1em}
\end{table}
We next propose a piecewise linear approximation of \eqref{eq:Lk_2} that is simpler to compute and, as shown in the next section, does not incur a performance penalty. LLRs with magnitude close to zero represent the least reliable bits which are critical for SD decoding. We therefore propose an approximation around $L_k=0$, i.e., the so-called zero-crossing (ZC) approximation introduced in \cite[Sec.~III-C]{alvarado2009distribution}. To derive expressions and compute the ZC points of $L_k$ and its slope at these points, we use the max-log approximation \cite[Eqs.~(4)-(6)]{viterbi1998intuitive} to simplify the summations inside the logarithm of \eqref{eq:Lk_2} to single exponential terms that are easier to handle analytically. For ease of notation, we now define $\sigma_1\triangleq \sigma(-3\Delta)$, $\sigma_2\triangleq \sigma(-\Delta)$, $\sigma_3\triangleq \sigma(\Delta)$, and $\sigma_4\triangleq \sigma(3\Delta)$, with $\mathrm{OMA}=3\Delta-(-3\Delta)=6\Delta$ being the optical modulation amplitude (OMA).

From Fig.~\ref{fig:BERR}(a), we see that $L_1$ has only one ZC point which is given by $-{\Delta(\sigma_3-\sigma_2)}/{(\sigma_3+\sigma_2)}$. The slope at this point is ${2\Delta}/{(\sigma_2\sigma_3)}$, and therefore the ZC approximation of $L_1^\mathrm{zc}$ is
\begin{equation}
    L_1^\mathrm{zc}=\frac{2\Delta}{\sigma_2\sigma_3}\left(y+\Delta\frac{\sigma_3-\sigma_2}{\sigma_3+\sigma_2}\right) \label{eq:L1_zc}.
\end{equation}
Next, we see from Fig.~\ref{fig:BERR}(a) that $L_2$ has two ZC points. The first is at $-{\Delta(3\sigma_2+\sigma_1)}/{(\sigma_2+\sigma_1)}$ with slope ${2\Delta}/{(\sigma_1\sigma_2)}$, while the second is at ${\Delta(\sigma_4+3\sigma_3)}/{(\sigma_4+\sigma_3)}$ with slope $-{2\Delta}/{(\sigma_3\sigma_4)}$. Since the slopes have opposite sign, the ZC approximation of $L_2^\mathrm{zc}$ is
\begin{align}
    L_2^\mathrm{zc}&=\min \left\{\frac{2\Delta}{\sigma_1\sigma_2}\left(y+\Delta\frac{3\sigma_2+\sigma_1}{\sigma_2+\sigma_1}\right),\right. \label{eq:L2_zc}\\
    &\hspace{36pt} \left.-\frac{2\Delta}{\sigma_3\sigma_4}\left(y-\Delta\frac{\sigma_4+3\sigma_3}{\sigma_4+\sigma_3}\right)\right\}. \notag
\end{align}
The ZC LLRs $\Lzc$ in \eqref{eq:L1_zc} and \eqref{eq:L2_zc} are shown with dashed lines in Fig.~\ref{fig:BERR}(a). LLR $L_1$ is approximated by a single line, while LLR $L_2$ is approximated by a piecewise linear function with two lines, since there are two ZC points. The ZC approximations are very close to the exact LLRs \eqref{eq:Lk_2} around the ZC points (black markers in Fig.~\ref{fig:BERR}(a)). In the next section, we analyze the performance of the SD FEC decoder using the three LLR calculations discussed, i.e., $L_k$, $\Lawgn$, and $\Lzc$.

\section{Numerical Results}

We use Monte Carlo simulations for the system in Fig.~\ref{fig:system} with the simulation parameters in Tab.~\ref{tab:sim_param}. We sweep the $\mathrm{OMA}$ from $-10$ to $5$ dBm, and calculate the post-FEC BER at every point. For SD decoding, we use the Chase decoding algorithm \cite{chase1972class} with $3$ least-reliable bits and flipping up to $2$ bits, resulting in $7$ test patterns. The results are shown in Fig.~\ref{fig:BERR}(b) in three groups of curves that represent net bitrates of $400$, $600$, and $800$ Gb/s. Note that the symbol rate, noise bandwidth, and constants $p_0$ and $p_1$ from Tab.~\ref{tab:sim_param} (for 400G) scale linearly with the bitrate. 

%Cross markers represent the post-FEC BER when using our proposed ZC approximation $\Lzc$. 
Figure~\ref{fig:BERR}(b) shows that for all cases the BER converges to an error floor at high OMA due to the presence of RIN in the channel. In general, the effect of RIN becomes more dominant as the data rate increases. Furthermore, there is a nonnegligible penalty caused by using the mismatched AWGN $\Lawgn$ instead of $L_k$ as we have predicted. At high OMA, using $\Lawgn$ results in an error floor increase of up to $5$ times for 400G, $3.3$ times for 600G, and $3$ times for 800G with respect to the exact $L_k$. Then, our low-complexity approximation $\Lzc$ recovers this penalty and matches the exact $L_k$ in terms of post-FEC BER. Therefore, we conclude that there is no performance loss when using our piecewise linear approximation $\Lzc$ instead of the quadratic functions resulting from the exact $L_k$, as the soft input for the SD decoder.

For the system parameters, this performance recovery also translates into achieving the KP4 FEC threshold of $2.26\times10^{-4}$ \cite{agrell2018} with $0.1$ and $0.4$ dB less in terms of OMA for 400G, and 600G, respectively. For 800G, the BER of $\Lawgn$ is already in the error floor region at the KP4 threshold. This represents a case where the mismatch in LLRs can lead to a scenario in which the KP4 threshold is not met or leaves no margin of operation. However, with $\Lzc$ we operate with reasonable margin, and thus, reliable error-free communication is feasible for next-generation transceivers.

%Lastly, we note that all the performance gains reported depend directly on the system parameters. In particular, as RIN decreases, the channel resembles an AWGN channel, and the performance penalty due to mismatched LLRs is reduced.

% \begin{figure}[!t]
%     \centering
%     \input{fig/pam4_sd_BW.tikz}
%     \caption{BW + FFE}
%     \label{fig:BER}
% \end{figure}

\section{Conclusions}
We studied the problem of LLR calculations for PAM-4 in next-generation RIN-dominated IM-DD systems. We showed that using mismatched AWGN LLRs results in a nonnegligible performance penalty, and proposed a low-complexity piecewise linear approximation that results in virtually no performance loss in terms of post-FEC BER. Future work includes extending the results to higher-order PAM, as well as experimental validation of the proposed technique with stronger (capacity-approaching) FEC codes.

\clearpage
\section{Acknowledgements}
This research is part of the project COmplexity-COnstrained LIght-coherent optical links \mbox{(COCOLI)} funded by Holland High Tech $|$ TKI HSTM via the PPS allowance scheme for public-private partnerships.
%-------------------------------------------------- Bibliography Section -------------------------------------------------------%
% see also https://tex.stackexchange.com/questions/55030/text-before-references-but-after-bibliography-title-with-bibtex as of 2024-02-29

\printbibliography[]

%%%%%%%%%%%%%%%%%%%%%%%%%%%%%%%%%%%%%%%%%%%%%
%---------------------------------------------- End of Document -----------------------------------------------%
\end{document}

%% file: fig/combined_system.tikz
\tikzstyle{block} = [draw, line width = 0.5pt, draw=gray,fill=gray!0, rectangle, minimum height=25pt, rounded corners=0.05cm, text width=2.4em,align=center, font=\footnotesize]
\tikzstyle{fecblock} = [draw, line width = 0.5pt, draw=gray,fill=gray!0, rectangle, minimum height=22pt, rounded corners=0.05cm, text width=3.4em,align=center, font=\footnotesize]
\tikzstyle{gain} = [draw=gray, line width=0.5pt, fill=gray!0, isosceles triangle, isosceles triangle apex angle=60, minimum height=22pt, minimum width=22pt, text width=1.8em, align=center, font=\footnotesize]

\tikzstyle{Cir} = [draw, circle,  minimum size=2.15em]
\tikzset{gaussian/.style={domain=-25:25,samples=100}}

%\definecolor{opt_block}{HTML}{6BAED6}  % light blue for optical blocks
%\definecolor{opt_block}{RGB}{251,106,74}
\definecolor{opt_block}{RGB}{128,128,255}
%\definecolor{fec_outer}{HTML}{969696}  % darker grey for KP4 Encoder / HD Decoder
%\definecolor{fec_inner}{HTML}{BDBDBD}  % medium grey for Hamming Encoder / SD Decoder
%\definecolor{fec_mod}{HTML}{D9D9D9}    % lighter grey for PAM-4 Mod / LLR Calc. D9D9D9
\definecolor{fec_outer}{RGB}{49,130,189}
\definecolor{fec_inner}{RGB}{107,174,214}
\definecolor{fec_mod}{RGB}{158,202,225}    
\definecolor{ch_color}{HTML}{38AA88}

\newcommand{\OColor}{blue!80!black}
\begin{tikzpicture}

% === Full system figure spanning two columns ===
\begin{scope}[scale=0.72, transform shape]
    % KP4 Encoder
    \node[fecblock, draw=fec_outer, fill=fec_outer!30](EncOut){KP4 Encoder};
    % Hamming Encoder
    \node[fecblock, below=0.8em of EncOut, draw=fec_inner, fill=fec_inner!30](EncIn){Hamming Encoder};
    % PAM-4 Mod
    \node[fecblock, below=0.8em of EncIn, draw=fec_mod, fill=fec_mod!30](Mod){PAM-4 Mod.};

    % === TX chain ===
    \node[block, right=1.8em of Mod, text width=1.3em,minimum height=4em] (tx_dsp) {\rotatebox{-90}{TX DSP}};
    \node[block, right=1.2em of tx_dsp] (DAC) {DAC};
    \node[gain, right=1.2em of DAC] (driver) {Driver};
    \node[block, right=1.2em of driver, draw=opt_block, fill=opt_block!0] (MZM) {MZM};
    \node[block, above=1.2em of MZM, text width=3.6em, draw=opt_block, fill=opt_block!0, minimum height=20pt, font=\footnotesize] (laser) {CW Laser};

    \node[red, left=2em of laser, font=\footnotesize](rin){RIN};
    \draw[draw,-stealth,red,densely dashed] (rin) -- (laser);

    % SMF symbol
    \draw [\OColor,solid,line width=0.4pt,opacity=1] ($(MZM.east)+(0.55cm,0.25)$) circle (2.5mm);
    \draw [\OColor,solid,line width=0.4pt,opacity=1] ($(MZM.east)+(0.65cm,0.25)$) circle (2.5mm);
    \draw [\OColor,solid,line width=0.4pt,opacity=1] ($(MZM.east)+(0.75cm,0.25)$) circle (2.5mm);
    \node[right=0.2cm of MZM, yshift=-0.6em, font=\footnotesize] (ssmf) {SMF};

    % === RX chain ===
    \node[block, right=1.3cm of MZM, draw=opt_block, fill=opt_block!0] (PD) {PD};
    \node[gain, right=1.0em of PD] (TIA) {TIA};
    \node[block, right=1.0em of TIA] (ADC) {ADC};
    \node[block, right=1.0em of ADC, text width=1.3em,minimum height=4em] (rx_dsp) {\rotatebox{-90}{RX DSP}};

    \node[red, above=1.8em of TIA, font=\footnotesize](th){Thermal Noise};
    \draw[draw,-stealth,red,densely dashed] (th) -- (TIA);

    % === FEC Decoder blocks ===
    \node[fecblock, right=1.8em of rx_dsp, draw=fec_mod, fill=fec_mod!30](LLR){LLR Calc.};
    \node[fecblock, above=0.8em of LLR, draw=fec_inner, fill=fec_inner!30](DecIn){SD Decoder};
    \node[fecblock, above=0.8em of DecIn, draw=fec_outer, fill=fec_outer!30](DecOut){KP4 Decoder};

    % BER measurement point - between SD Decoder and HD Decoder (vertical)
    \coordinate (bermid) at ($(DecIn.north)!0.33!(DecOut.south)$);
    \node[right=2.7em of bermid, inner sep=1pt,font=\small](ber){$\mathrm{BER}$};
    \coordinate (llrmid) at ($(LLR.north)!0.33!(DecIn.south)$);
    \node[right=2.7em of llrmid, inner sep=1pt,font=\small](Lk){$L_k$};
    
    % === Arrows ===
    % Encoder chain
    \draw[draw,-stealth] (EncOut) -- (EncIn);
    \draw[draw,-stealth] (EncIn) -- (Mod);
    \draw[draw,-stealth] (Mod) -- node[above,font=\small,xshift=-0.2em]{$X$}(tx_dsp);

    % TX chain
    \draw[draw,-stealth] (tx_dsp) -- (DAC);
    \draw[draw,-stealth] (DAC) -- (driver);
    \draw[draw,-stealth] (driver) -- (MZM);
    \draw[draw,-stealth,\OColor] (laser) -- (MZM);
    \draw[draw,-stealth,\OColor] (MZM) -- (PD);

    % RX chain
    \draw[draw,-stealth] (PD) -- (TIA);
    \draw[draw,-stealth] (TIA) -- (ADC);
    \draw[draw,-stealth] (ADC) -- (rx_dsp);

    % Decoder chain
    \draw[draw,-stealth] (rx_dsp) -- node[above,font=\small,xshift=0.2em]{$Y$}(LLR);
    \draw[draw,-stealth] (LLR) -- node[above,font=\small]{}(DecIn);
    \draw[draw,-stealth] (DecIn) -- (DecOut);
    \draw[draw,-stealth,densely dotted,gray] (bermid) -- (ber);
    \draw[draw,-stealth,densely dotted,gray] (llrmid) -- (Lk);

    % Legend - below KP4, Hamming, PAM-4 blocks
    \node[right=1.3em of EncOut, yshift=-1em] (legend0) {};
    \draw[draw,-stealth] ($(legend0)$) -- node[right,xshift=0.6em,font=\footnotesize]{Electrical}($(legend0)+(1.8em,0)$);
    \draw[draw,-stealth,\OColor] ($(legend0)+(0em,-1em)$) -- node[right,xshift=0.6em,yshift=-1pt,font=\footnotesize]{\textcolor{black}{Optical}}($(legend0)+(1.8em,-1em)$);

    % Channel model equation
    \node[below=1.2em of PD, xshift=-2.5em, font=\footnotesize] (ch_eq) {Eq.~\eqref{eq:ch_model} : \normalsize{$Y=X+Z\sigma(X)$}};
    \draw[draw,-stealth,densely dotted,line width=0.5pt, color=gray] ($(Mod.east)+(0.7em,0)$) |- (ch_eq);
    \draw[draw,-stealth,densely dotted,line width=0.5pt, color=gray] ($(rx_dsp.east)+(0.9em,0)$) |- (ch_eq);

    \begin{pgfonlayer}{background}
        \fill[fill=ch_color!10, rounded corners=0.05cm] ($(Mod.east)+(1.3em,6em)$) rectangle ($(LLR.west)+(-1.4em,-2.6em)$);
        \draw[rounded corners=0.05cm, densely dashed, ch_color!50, line width=0.4pt] ($(Mod.east)+(1.3em,6em)$) rectangle ($(LLR.west)+(-1.4em,-2.6em)$);
        \node[above=8.2em of ch_eq,font=\footnotesize](imdd){IM-DD System};
        %\node[above=2.2em of RS_in] () {Host};
    \end{pgfonlayer}

    % === Eye diagram (below LLR Calc. block, pushed lower) ===
    \definecolor{matlab1}{RGB}{252,174,145}
    \definecolor{matlab2}{RGB}{251,106,74}
    \definecolor{matlab3}{RGB}{222,45,38}
    \definecolor{matlab4}{RGB}{165,15,21}

    \def\xA{81};
    \def\xB{56};
    \def\xC{32};
    \def\xD{7};
    \def\colorA{matlab1};
    \def\colorB{matlab2};
    \def\colorC{matlab3};
    \def\colorD{matlab4};
    \def\colorOpac{0.35};

    % Position the eye diagram below the LLR Calc. block
    \node[right=4em of LLR, yshift=-3.5em] (eye){};
    %\draw[draw,-stealth,dashed,red] ($(ADC.east)+(0.4em,0)$) |- ($(eye)+(0em,4em)$);
    \draw[draw,-stealth,densely dashed,red] ($(ADC.east)+(0.4em,0)$) |- ($(eye)+(-0.2em,1.2em)$);

    \begin{scope}[shift={(eye)}]
    \begin{axis}[
        width=2.1in,
        height=2.1in,
        xmin=1, xmax=99,
        ymin=0, ymax=100,
        enlargelimits=false,
        ylabel style={yshift=-8pt},
        xlabel style={yshift=2pt},
        ylabel={PAM-4 Eye Diagram},
        ytick={\xD,19,\xC,44,\xB,68.5,\xA,93.5},
        xtick={12.5,25,...,100},
        yticklabels=\empty,
        xticklabels=\empty,
        font=\footnotesize,
        grid=both,
        grid style={dashed,lightgray!75},
        ]
        \addplot[] graphics[xmin=0,ymin=2,xmax=100,ymax=102] {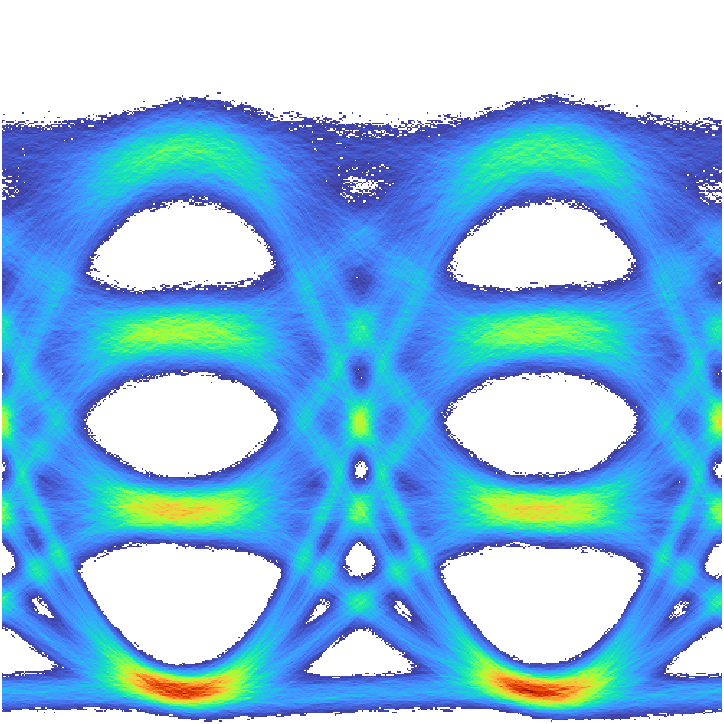};
        \draw[black, dashed, line width=0.5] (25,0) -- (25,90);

        \node[black, font=\footnotesize] at (25,94) {$Y$};
        \draw[draw=none,fill=white,opacity=0.85] (axis cs:75,0) rectangle (100,100);

        \def\sA{6.0};
        \def\sB{4.8};
        \def\sC{3.64};
        \def\sD{2.63};
        \def\scale{85};

        \draw[gaussian, \colorA, fill=\colorA, fill opacity=\colorOpac]
        plot[domain=-20:20]({75 + \scale/sqrt(2*pi*\sA)*exp(-(\x)^2/(2*\sA^2))}, {\xA + \x});
        \draw[gaussian, \colorB, fill=\colorB, fill opacity=\colorOpac]
        plot[domain=-15:15]({75 + \scale/sqrt(2*pi*\sB)*exp(-(\x)^2/(2*\sB^2))}, {\xB + \x});
        \draw[gaussian, \colorC, fill=\colorC, fill opacity=\colorOpac]
        plot[domain=-15:15]({75 + \scale/sqrt(2*pi*\sC)*exp(-(\x)^2/(2*\sC^2))}, {\xC + \x});
        \draw[gaussian, \colorD, fill=\colorD, fill opacity=\colorOpac]
        plot[domain=-15:15]({75 + \scale/sqrt(2*pi*\sD)*exp(-(\x)^2/(2*\sD^2))}, {\xD + \x});
        \draw[black, densely dashed, line width=0.7] (75,0) -- (75,100);

    \end{axis}
    \end{scope}

\end{scope}

\end{tikzpicture}

%% file: fig/pam4_combined.tex
% 0: OMA | 1: LLR_exact_k1 | 2: LLR_exact_k2 | 3: LLR_awgn_k1 | 4: LLR_awgn_k2
\pgfplotstableread{data_txt/LLR_PAM4_OMA-0_RIN-140.txt}\dataLLR
\pgfplotstableread{data_txt/BER_PAM4_RIN-140_400G.txt}\dataPAMFOUR
\pgfplotstableread{data_txt/BER_PAM4_RIN-140_800G.txt}\dataPAMEIGHT
\pgfplotstableread{data_txt/BER_PAM4_RIN-140_600G.txt}\dataPAMSIX
\def\lw{0.6pt}
\def\mklw{0.5pt}
\def\mksz{0.6}
\def\mkcross{1.2}
\def\mkrr{x}

\definecolor{col400}{RGB}{35,139,69}
\definecolor{col600}{RGB}{65,171,93}
\definecolor{col800}{RGB}{116,196,118}

\pgfplotsset{
    curveLk/.style={myGreen, line width=\lw, mark=none, mark options={scale=1*\mksz,line width=\mklw,fill=white},mark layer=like plot}
}
\pgfplotsset{
    curveLawgn/.style={myBlue, line width=\lw, mark=none, mark options={scale=0.8*\mksz,line width=\mklw,fill=white},mark layer=like plot}
}
\pgfplotsset{
    curveLzc/.style={myRed, dashed, line width=\lw, mark=*, mark options={scale=\mksz,line width=\mklw,fill=white,solid},mark layer=like plot}
}

\begin{tikzpicture}[scale=1]
    \begin{axis}[
        hide axis,
        xmin=0, xmax=1, ymin=0, ymax=1,
        legend columns=-1,
        legend style={draw=black,at={(0.5,0.5)},anchor=center,column sep=1ex,font=\footnotesize},
    ]
        \addlegendimage{curveLk};
        \addlegendimage{curveLawgn};
        \addlegendimage{curveLzc,densely dashed};
        
        \addlegendentry{Exact $L_k$ \eqref{eq:Lk_2}};
        \addlegendentry{AWGN $\Lawgn$ \eqref{eq:Lk_awgn}};
        \addlegendentry{ZC Approx. $\Lzc$ \eqref{eq:L1_zc} and \eqref{eq:L2_zc}};
    \end{axis}
\end{tikzpicture}

\begin{tikzpicture}[scale=1]
    \begin{groupplot}[
        group style={
            group size=2 by 1,
            ylabels at=edge left,
            xlabels at=edge bottom,
            vertical sep=0.75cm,
            horizontal sep=0.1cm,
        },
        width=0.7\columnwidth,
        height=2.4in,
        xmin=-3.5, xmax=3.5,
        xlabel={Channel Output $y$},
        xtick={-3,-2,...,3},
        xticklabels={$-3\Delta$,,$-\Delta$,,$\Delta$,,$3\Delta$},
        ytick={-200,-100,...,200},
        xlabel style={yshift=4pt},
        ylabel style={yshift=-8pt},
        title style={yshift=-20pt,xshift=0pt},
        xtick pos=bottom,
        ymin=-200, ymax=200,
        ylabel={LLR},
        grid=both,
        grid style = {dashed,lightgray!50,line width=0.3pt},
        legend columns=-1,
        legend style = {at={(0.85,1.1)}, anchor=south, font=\scriptsize, legend cell align=left, row sep=-0.5ex, column sep=0.5ex,inner sep=0.2ex},
        %legend to name=sharedlegend,
        %font=\footnotesize,
        every axis/.append style={font=\footnotesize},
    ]

    \node[font=\normalsize] at (-0.35in, 1.7in) {(a)};
    % --- LLR ---
    \nextgroupplot[
        title={Bit Position $k=1$},
    ]
    \addplot[myGreen, line width=\lw, mark=none, mark options={scale=1.4*\mksz,line width=\mklw,fill=white}] table[x index=0, y index=3]{\dataLLR};
    \addplot[myBlue, solid, line width=\lw, mark=none, mark options={scale=1*\mksz,solid,line width=\mklw,fill=white}] table[x index=0, y index=1]{\dataLLR};    
    \addplot[myRed, densely dashed, line width=\lw, mark=none, mark repeat=12, mark phase=12,mark options={scale=0.8*\mksz,line width=\mklw,fill=white}] table[x index=0, y index=5]{\dataLLR};

    %\addlegendentry{Max-log \eqref{eq:Lk_2}};
    %\addlegendentry{AWGN \eqref{eq:Lk_awgn}};
    %\addlegendentry{ZC Approx.\ \eqref{eq:L1_zc}+\eqref{eq:L2_zc}};

    \addplot[black,densely dotted] coordinates{(-4,0) (-0.1588, 0) (-0.1588, -200)};
    \addplot[myRed,only marks,mark=*,mark options={scale=\mksz,line width=\mklw,fill=white}] coordinates{(-0.1588,0)};
    \node[myRed,anchor=north west,font=\scriptsize] at (axis cs:-0.1588,0){$\mathrm{ZC}$};
    %\node[font=\scriptsize, anchor=south east, inner sep=1pt] at (rel axis cs:0.99,0.01)(zc){$\mathrm{ZC}:(x_1^0,x_1^1)=(-\Delta,\Delta)$};
    %\draw[draw,-stealth,black!70,densely dotted] (axis cs:-0.1588,0) -- (-1.8,-160);

    \nextgroupplot[
        yticklabels=\empty,
        title={Bit Position $k=2$},
    ]
    \addplot[myGreen, line width=\lw, mark=none, mark options={scale=1.4*\mksz,line width=\mklw,fill=white}] table[x index=0, y index=4]{\dataLLR};
    \addplot[myBlue, solid, line width=\lw, mark=none, mark options={scale=1*\mksz,solid,line width=\mklw,fill=white}] table[x index=0, y index=2]{\dataLLR};
    \addplot[myRed, densely dashed, line width=\lw, mark=none, mark repeat=12, mark phase=12, mark options={scale=0.8*\mksz,line width=\mklw,fill=white}] table[x index=0, y index=6]{\dataLLR};

    \addplot[black,densely dotted] coordinates{(-4,0) (-2.2117, 0) (-2.2117, -200)};
    \addplot[myRed,only marks,mark=*,mark options={scale=\mksz,line width=\mklw,fill=white}] coordinates{(-2.2117,0)};
    \node[myRed,anchor=north west,font=\scriptsize] at (axis cs:-2.2117,0){$\mathrm{ZC}_1$};
    %\node[font=\scriptsize, anchor=north west, inner sep=1pt] at (rel axis cs:0.01,0.85)(zc1){$\mathrm{ZC}_1:(x_2^0,x_2^1)=(-3\Delta,-\Delta)$};
    %\draw[draw,-stealth,black!70,densely dotted] (axis cs:-2.2117,0) -- (-2.9,110);

    \addplot[black,densely dotted] coordinates{(-4,0) (1.8871, 0) (1.8871, -200)};
    \addplot[myRed,only marks,mark=*,mark options={scale=\mksz,line width=\mklw,fill=white}] coordinates{(1.8871,0)};
    \node[myRed,anchor=north east,font=\scriptsize] at (axis cs:1.8871,0){$\mathrm{ZC}_2$};
    %\node[font=\scriptsize, anchor=south east, inner sep=1pt] at (rel axis cs:0.99,0.01)(zc2){$\mathrm{ZC}_2:(x_2^0,x_2^1)=(3\Delta,\Delta)$};
    %\draw[draw,-stealth,black!70,densely dotted] (axis cs:1.8871,0) -- (-1.9,-160);
    
    \end{groupplot}

    %\node at ($(group c1r1.north)!0.5!(group c2r1.north)+(0,1cm)$)
    %{\ref{sharedlegend}};
    
\end{tikzpicture}
~%\hspace{-0.5em}
\begin{tikzpicture}[scale=1]
    \begin{semilogyaxis}[
        width=0.9*\columnwidth,
        height=2.4in,
        xmin=-10, xmax=5, 
        ymin=5e-9, ymax=1e-1,
        xlabel={OMA [dBm]},
        xtick={-14,-12,...,10},
        ytickten={-9,-8,...,1},
        ylabel={Post-FEC BER},
        ylabel style={yshift=0pt},
        xlabel style={yshift=4pt},
        grid=major,
        grid style = {dashed,lightgray!50},
        legend columns=-1,
        legend style = {at={(0.35,1.02)}, anchor=south, font=\scriptsize, legend cell align=left, row sep=-0.5ex, column sep=0.5ex,inner sep=0.2ex},
        font=\footnotesize,
        set layers=standard,
        %every axis/.append style={font=\footnotesize},
    ]
    
    % --- BER ---
    % ---------- 400G ---------
    \addplot[curveLk] table[x index=0, y index=1]{\dataPAMFOUR};
    \addplot[curveLawgn] table[x index=0, y index=3]{\dataPAMFOUR};
    \addplot[curveLzc] table[x index=0, y index=2]{\dataPAMFOUR};

    % ---------- 600G ---------
    \addplot[curveLk] table[x index=0, y index=1]{\dataPAMSIX};
    \addplot[curveLawgn] table[x index=0, y index=3]{\dataPAMSIX};
    \addplot[curveLzc] table[x index=0, y index=2]{\dataPAMSIX};

    % ---------- 800G ---------
    \addplot[curveLk] table[x index=0, y index=1]{\dataPAMEIGHT};
    \addplot[curveLawgn] table[x index=0, y index=3]{\dataPAMEIGHT};
    \addplot[curveLzc] table[x index=0, y index=2]{\dataPAMEIGHT};

    \draw[fill=white,draw=none,opacity=0.7] (axis cs:-9.9, 9e-5) rectangle (-7.2,2e-4);
    \node[font=\scriptsize,anchor=south,black!70] at (axis cs:-8.4, 6e-5)(){KP4-FEC};
    \addplot[gray,line width=0.8pt] coordinates{(-15,2.26e-4) (10,2.26e-4)};
    \draw[line width = 0.4pt, col400] (axis cs:-1,3e-8) ellipse (2mm and 4mm);
    \draw[line width = 0.4pt, col600] (axis cs:1,6e-6) ellipse (2mm and 4mm);
    \draw[line width = 0.4pt, col800] (axis cs:2,1.5e-4) ellipse (2mm and 4mm);
    \node[font=\scriptsize,anchor=south, fill=white, draw=col400, inner sep=2pt, rounded corners=0.7mm, line width=0.5pt] at (axis cs:-1.2,1.5e-7)(){400G};
    \node[font=\scriptsize,anchor=north, fill=white, draw=col600, inner sep=2pt, rounded corners=0.7mm, line width=0.5pt] at (axis cs:1.8,1.2e-6)(){600G};
    \node[font=\scriptsize,anchor=south, fill=white, draw=col800, inner sep=2pt, rounded corners=0.7mm, line width=0.5pt] at (axis cs:2.6,7.5e-4)(){800G};

    % TODO (Felipe): "IN THE OVEN" — update simulation results when ready
    %\node[red] at (axis cs:-7,1e-8)(asd){IN THE OVEN};
    %\draw[draw,-stealth,red,line width=2pt] (asd) -- (axis cs:0, 1e-8);

    % TODO (Felipe): Add double-headed arrows at high OMA showing error floor increase factors (5x for 400G, 3x for 800G)
    % between the max-log and AWGN curves once simulation data is finalized.
    \def\arrowCol{black!70}
    \draw[draw,-stealth,\arrowCol] (axis cs:3.5, 9e-5) -- (3.5, 2.26e-4);
    \draw[draw,-stealth,\arrowCol] (axis cs:3.5, 2.9e-6) -- (3.5, 1e-5);
    \draw[draw,-stealth,\arrowCol] (axis cs:3.5, 8e-9) -- (3.5, 4.5e-8);

    \node[\arrowCol,anchor=south west,font=\scriptsize] at (axis cs:2.8, 2e-5){$\times 3$}; 
    \node[\arrowCol, anchor=south west,font=\scriptsize] at (axis cs:2.8, 6e-7){$\times 3.3$};
    \node[\arrowCol, anchor=south west,font=\scriptsize] at (axis cs:2.8, 4e-8){$\times 5$}; 
    
    \end{semilogyaxis}

    \node[font=\normalsize] at (-0.45in, 1.7in) {(b)};
    
\end{tikzpicture}